\documentclass[%
 aip,
 jcp,
 amsmath,amssymb,
 reprint,%
]{revtex4-1}

\usepackage{graphicx}
\usepackage{dcolumn}
\usepackage{bm}

\usepackage[utf8]{inputenc}
\usepackage[T1]{fontenc}
\usepackage{mathptmx}

\usepackage{hyperref}
\usepackage{color}

\newcommand{\angstrom}{\textup{\AA}}

\begin{document}

\preprint{AIP/123-QED}

\title{Entropy of different phases formed by soft rods}

\author{Jayeeta Chattopadhyay}
\affiliation{%
Centre for Condensed Matter Theory, Department of Physics, Indian Institute of Science, Bangalore 560012, India
}%

 \author{Shiang-Tai Lin}%
\affiliation{Department of Chemical Engineering, National Taiwan University, Taipei 10617, Taiwan}

\author{Prabal K. Maiti}
\email{maiti@iisc.ac.in}
\affiliation{%
Centre for Condensed Matter Theory, Department of Physics, Indian Institute of Science, Bangalore 560012, India
}%
\date{\today}

\begin{abstract}
Computation of entropy in liquids and liquid crystal phases is a big challenge in statistical physics. In this work, we extend the two-phase thermodynamic model (2PT) to shape anisotropic  soft repulsive spherocylinders (SRSs) and report the absolute values of entropy for different liquid crystal (LC) phases for a range of aspect ratios $L/D = 2-5$. We calculate the density of states (DoS) for different LC phases and decompose it into contributions arising from translational and rotational degrees of freedom. 
The translational and rotational modes are further partitioned into diffusive, gas-like, and non-diffusive, solid-like components using a fluidicity factor. 
In the dilute limit, the entropy values obtained from the 2PT method match exactly those of an ideal rigid rotor. We find that, for a given packing fraction, the magnitude of the total entropy is roughly equal regardless of the different LC phases associated with different aspect ratios. We also compute the excess entropy (for $L/D = 5$) and compare those with the values obtained using the standard integration approach of molecular dynamics (MD) or Monte Carlo (MC) equation of state (EOS) of SRS. The values obtained using both approaches match very well. The rotational and translational fluidicity factors are further used to determine the phase boundaries of different liquid crystal phases for the respective aspect ratios.
\end{abstract}

\maketitle

\section{\label{sec:level1} Introduction}
The phase behavior of shape anisotropic particles is an emerging field of research that gives rise to various liquid crystal (LC) phases \cite{de-gennes,bolhuis,McGrother}. Examples span from living organisms like tobacco mosaic virus \cite{Dogic-PRL-TMV,Lowen-PRE-TMV,TMV-PRL-1989}, fd virus \cite{Dogic-PRL-FD} to synthetic systems of rod-like particles like boehmite \cite{boehmite-JPC-1993}, silica \cite{silica-2012} etc. Different liquid crystal phases can be identified based on their microscopic arrangements, as well as positional and orientational order.

Onsager, in his seminal work \cite{onsager}, showed that a system of thin and hard rods could undergo a phase transition from disordered isotropic to orientationally ordered nematic phase above a critical aspect ratio ($L/D > 3.7$) that is mainly driven by entropy.
The loss of orientational entropy in the nematic phase is compensated by the increase of translational entropy due to the ordered structure. Similarly, for the other LC phases, entropy plays an important role in studying the stability of the phases. Entropy of a fluid can be expressed as a multiparticle correlation expansion of statistical entropy developed by Green and Nettleton \cite{green1952molecular,nettleton1958expression} and generalized by Lazaridis and co-workers \cite{lazaridis1992entropy,lazaridis1996orientational} for the non-spherical bodies. Costa \textit{et al.} first used this method to calculate the entropy of a system of hard spherocylinders (HSCs) \cite{costa1998cpl, costa2002entropy} and later, by Cuetos \textit{et al.} \cite{cuetos-2002} in a system of soft repulsive spherocylinders (SRSs). It is also worth mentioning several interesting works by Dhar \textit{et al.} \cite{dhar-EPL-2007,Dhar-PRE-2021} where they have calculated entropy of hard rods and rigid rectangles in 3D and 2D using analytically solvable lattice model and MC simulations. 

In 2003, Lin \textit{et al.} \cite{lin2003two} developed the two-phase thermodynamic (2PT) model to calculate the entropy, free energy, and other thermodynamic properties of liquids from a short MD trajectory (20 picoseconds (ps)). 2PT model has emerged as an efficient and accurate method in calculating various thermodynamic properties of Lennard-Jones fluids for the diverse setting of state points both in 2D \cite{pannir2018efficient} and 3D \cite{lin2003two}, water in bulk \cite{H20-2PT} and under different confinement, carbon dioxide \cite{CO2-2PT} and other organic and inorganic molecules \cite{borah2012transport}. The results match very well with those of the experimental studies. In the 2PT method, the density of state (DoS) of a liquid, which is calculated from the Fourier transform of the velocity autocorrelation function (VACF), is decomposed into vibrational (solid) and diffusive (gas) components. The thermodynamic quantities, including entropy, are then calculated using harmonic oscillator approximation to the solid component and hard sphere approximation to the gas component. For the rotational mode, the diffusive part is calculated from the rigid rotor approximation \cite{lin2003two,H20-2PT}. In 2PT method, the entropy of a definite state point is calculated from a single MD trajectory. Thus, it is far more efficient than the conventional integration approach of MD or MC equation of state of the SRS, which entails several discrete MD/MC trajectories along the integration path. This is advantageous for the systems for which the analytical form of the equation of state is unknown (such as SRS).

In this work, we extend the 2PT method to calculate entropy of various liquid crystal phases formed by a system of soft repulsive spherocylinders of different aspect ratios (length/diameter) $L/D = 2, 3, 3.5, 4 $ and 5. 
We validate our method by comparing the entropy values  obtained using the standard integration approach of equation of state of the SRS of $L/D = 5$ at $T^{*} = 5$ \cite{cuetos-2002, costa2002entropy}. We find that the entropy values do not have any strong dependence on the aspect ratio but strongly depend on the packing fraction ($\eta$)of the system. We also find that LC phase transitions are governed by the change of pair entropy. The loss of orientational pair entropy in the nematic phase is compensated by the increase of translational pair entropy. Similarly, in case of the smectic phase, the loss of translational pair entropy is compensated by the residual entropy arising from the multi-particle contribution.   
In addition, we present an alternative way to identify the phase boundaries of different liquid crystal phases from the fluidicity factor that is directly related to the diffusivity of the system: the packing fraction at which the translational fluidicity $f_{trans}$ saturates but rotational fluidicity $f_{rot}$ decreases sharply indicates the phase boundary of the isotropic to nematic (I-N) phase transition. Similarly, the nematic to smectic (N-Sm) transition is located where $f_{rot}$ saturates but $f_{trans}$ keeps decreasing. 

The rest of the paper is organized as follows: In section II, we briefly describe the theoretical background of the 2PT method and summarize the multiparticle correlation expansions method and the integration approach of equation of state to calculate the entropy of SRS; in section III, we describe the SRS model  and the simulation protocol. We present the results and analysis in section IV. Finally, in section V, we conclude with the discussion on the major benefits of the 2PT method and possible applications.

\section{\label{sec:level2} Model and Computational Details}
We model the system as a collection of spherocylinders (cylinder with hemispherical caps) of aspect ratios $L/D = 2, 3, 3.5,4, 5 $.
The interacting potential is only due to the excluded volume interaction described by the Weeks-Chandler-Andersen (WCA) potential given as follows \cite{weeks1971}: 


\begin{equation}
\begin{split}
U_{SRS} & =  4\varepsilon[(\dfrac{D}{d_{m}})^{12}-(\dfrac{D}{d_{m}})^{6}]+\epsilon \hspace{0.8cm} if \hspace{0.3cm} d_{m} < 2^{\frac{1}{6}}D\\
& = 0 \hspace{4.0cm} if \hspace{0.3cm} d_{m} \geq 2^{\frac{1}{6}}D
\end{split}
\end{equation} 

Here, $d_{m}$ is the shortest distance between two SRS that determines their relative orientation \cite{allen1993hard,vega1994fast,bolhuis,cuetos-2002,jaydeep-softmatter-2022}. 
For convenience, thermodynamic quantities are expressed in terms of interaction strength  $\epsilon$, diameter of the SRS $D$ and mass $m$:
temperature $T^{*}=\frac{k_{B}T}{\epsilon} $, pressure $ P^{*}= \dfrac{Pv_{hsc}}{k_{B}T} $, packing fraction $ \eta = v_{hsc}\rho$, where $\rho$ is the number density of the system defined as, $ \rho = \frac{N}{V} $ and $ v_{hsc}= \pi D^{2}(\frac{D}{6} + \frac{L}{4})$ is the volume of the spherocylinder; energy $ E^{*} = \frac{E}{\epsilon} $, entropy $ S^{*} = \frac{S}{k_{B}}$
, Helmholtz free energy $ A^{*} = \frac{A}{\epsilon} $, Gibbs free energy $ G^{*} = \frac{G}{\epsilon} $, diffusivity $ d^{*} = d(\frac{m}{\epsilon})^{1/2}/D $ and the time $t^{*} = t \sqrt{\epsilon/m}/D$.
To compute entropy using 2PT method, we convert all the thermodynamic quantities in real units using the parameters of argon ($\epsilon = 0.238$  kcal/mol, $\sigma = 3.405 \angstrom $ and mass $m = 39.948$ g/mol) and then again convert them into the reduced units.\\

We build the system in a hexagonal-closed-packed (HCP) crystal structure. As the particles are inherently anisotropic in shape, we choose the number of particles in the x, y and z directions such that the simulation box can be built in a near-cubic geometry. If $n_{x}$, $n_{y}$, $n_{z}$ are the number of particles in the x, y, z direction respectively and $n_{u}$ is the number of particles in one unit cell, then the total number of particles in one simulation box $N = n_{u} \times n_{x} \times n_{y} \times n_{z}$. In our case, number of SRSs is chosen to be $N = 1024$.
The  periodic boundary condition in all three directions are used. \\

 We have carried out a series of MD simulations for a wide range of state points spanning the melting transition from solid (crystal) to gas (isotropic) for all the  aspect ratios. We melt the initial crystal structure slowly by reducing the pressure in NPT ensemble (Constant particle number, pressure and temperature) at $T^{*} = 5 $ for each $L/D$.  The positions and velocities of the SRSs are updated using Verlet algorithm \cite{verlet1967} and the rotational motion by quaternion-based rigid-body dynamics \cite{omelyan1998,martys1999,rotunno2004,lansac2003,maiti2002}. The temperature and pressure of the system are controlled using Berendsen thermostat and barostat \cite{berendsen1984} with a temperature relaxation time $\tau_{T} = 0.05 $ and pressure relaxation time $\tau_{P} = 2 $ respectively. We perform $1 \times 10^{5}$ to $2 \times 10^{5}$ MD steps (with an integration time step $\delta t = 0.001$ in reduced unit) to reach equilibrium condition and another $2-5 \times 10^{3} $ steps (5-30 ps in real unit using the above-mentioned parameters) with $\delta t = 5 \times 10^{-4} $(1 fs)in real unit for the 2PT method. 


\section{\label{sec:level2} Theory}
    \subsection{Two phase thermodynamic method}
        \subsubsection{Density of State Function}

The density of state (DoS) function $G(\nu )$ is defined as the mass weighted sum of the atomic spectral densities. This can be obtained from Fourier transform of velocity auto-correlation function (VACF) obtained from MD trajectory \cite{lin2003two}.

\begin{equation}
    G(\nu) = \frac{1}{k_{B}T} \sum_{l=1}^{N_{atom}}\sum_{k=1}^{3}\lim_{\tau \to -\infty} \frac{m_{l}}{\tau}{\displaystyle\left \lvert\int_{-\tau}^{\tau}v_{l}^{k}(t) e^{-i2\pi\nu t}dt \right \rvert}^{2}
\end{equation}

Here, $N_{atom}$ is the total number of atoms in the system. $m_{l}$ is  mass of the $l^{th}$ atom and $v_{l}^{k}$ is the velocity of the $l^{th}$ atom in $k^{th}$ direction ( $k$ indicates spatial coordinates $x, y, z$ respectively). $G(\nu)$ represents distribution of normal modes in the system i.e $G(\nu)$ $d\nu$ represents number of normal modes in the frequency range  $\nu$ to $\nu + d\nu$. So, total number of modes in the system i.e degrees of freedom of the system $3N$
\begin{equation}
\int_{0}^{\infty} G(\nu) d\nu = 3N
\end{equation}

The diffusion constant ($D$) of the system is directly related to the zero-frequency density of state of the system $G(0)$:
\begin{equation}
    D = \dfrac{k_{B}T}{12mN}G(0)
\end{equation}

For a rigid SRS, there is no vibrational motion. So, total number of degrees of freedom for a rigid SRS is 5 comprising 3 translational and 2 rotational motion. Therefore, total number of modes in the system is:
\begin{equation}
\int_{0}^{\infty} G(\nu) d\nu = 5N
\end{equation}

Density of state $G(\nu)$ is decomposed into translational and rotational part:
\begin{equation}
    G(\nu) = G_{trans}(\nu) + G_{rot}(\nu)
\end{equation}
where, $G_{trans}(\nu)$ is obtained from the translational component of the center of mass velocity of the SRS:

\begin{equation}
    G_{trans}(\nu) = \frac{1}{k_{B}T} \sum_{j=1}^{N}\sum_{k=1}^{3}\lim_{\tau \to -\infty} \frac{m_{j}}{\tau}{\displaystyle\left \lvert\int_{-\tau}^{\tau}v_{j}^{k^{trans}}(t) e^{-i2\pi\nu t}dt \right \rvert}^{2}
\end{equation}

here, $N$ is the total number of SRS in the system and $m_j$ is the mass of the SRS. $v_{j}^{k^{trans}}$ is the translational velocity of $j^{th}$ SRS in $k^{th}$ direction.

\begin{equation}
    G_{rot}(\nu) = \frac{1}{k_{B}T} \sum_{j=1}^{N}\sum_{k=1}^{2}\lim_{\tau \to -\infty} \frac{I_{j}^{k}}{\tau}{\displaystyle\left \lvert\int_{-\tau}^{\tau}\omega_{j}^{k}(t) e^{-i2\pi\nu t}dt \right \rvert}^{2}
\end{equation}

here, $I_{j}^{k}$ is the moment of inertia of $j^{th}$ SRS along $k^{th}$ the principal axis. As SRS is linear, the moment of inertia along its director is 0. Therefore, $k$ runs from 1 to 2. $\omega_{j}^{k}$ represents the angular velocity.


\subsubsection{Thermodynamic properties from 2PT method}

Various thermodynamic quantities like energy, entropy of a system can be expressed as a summation over the 
contributions from translational and rotational  motion of SRS) \cite{H20-2PT,CO2-2PT}:
\begin{equation}
E = E_{0} + E_{trans} +  E_{rot} , \label{eq-energy}
\end{equation}
\begin{equation}
S = S_{trans} +  S_{rot} . \label{eq-entropy}
\end{equation}Here, $E_{0}$ is the reference energy. In 2PT method, the density of states corresponding to translational or rotational 
motion is partitioned  as:
\begin{equation}
G_{k}(\nu) = G_{k}^{s}(\nu) +  G_{k}^{g}(\nu)
\label{eq-dos}
\end{equation}
where, the subscript k stands for  translational, or rotational motion. The 1st term in Eq. \ref{eq-dos} refers to the solid-like and the  2nd term in Eq. \ref{eq-dos} refers to the gas-like contributions. For a solid-like system, the DoS can be exactly determined by that of harmonic oscillator. But for a liquid, harmonic approximation is no longer valid at the low frequency regime due to the strong effect of anharmonicity. Also, the diffusive model at the zero frequency can lead to singularity. In the 2PT model, the anharmonicity effect at the low frequency is treated by decomposing the DoS into gas-like and solid-like components as mentioned in Eq. \ref{eq-dos}. The gas-like component is evaluated from the DoS at the zero frequency and the fluidicity factor $f_{k}$ using the following equation :
\begin{equation}
G_{k}^{g}(\nu) = \frac{G_{k}(0)}{1+\left[\frac{\pi\nu G_{k}(0)}{6f_{k}N}\right]^{2}} . \label{eq-dos-gas}
\end{equation}The fluidicity factor $f_{k}$ is calculated using the equation below:
\begin{equation}
2 \Delta_{k}^{-9/2} f_{k}^{15/2} - 6\Delta_{k}^{-3}f_{k}^{5} - \Delta_{k}^{-3/2} f_{k}^{7/2} + 6\Delta_{k}^{-3/2} f_{k}^{5/2} + 2f_{k} - 2 = 0 ,
\end{equation}where, $\Delta_{k}$ is the diffusivity constant in reduced unit that is defined as:
\begin{equation}
\Delta_{k}(T, V, N, k, G_{k}(0)) = \frac{2 G_{k}(0)}{9N} \left(\frac{\pi k_{B}T}{k}\right)^{1/2} \left(\frac{N}{V}\right)^{1/3} \left(\frac{6}{\pi}\right)^{2/3} . \label{eq-diffusivity}
\end{equation}The above equation Eq. \ref{eq-diffusivity} indicates $\Delta_{k}$ only depends on the thermodynamic state points ($T, V, N$) and $G_{k}(0)$ that can uniquely determines the fluidicity factor $f_{k}$ for different modes. Once we calculate $G_{k}^{g}(\nu)$ from Eq. \ref{eq-dos-gas}, the solid-like component can be determined by subtracting it from the total DoS $G_{k}(\nu)$ (Eq. \ref{eq-dos}) obtained  from velocity auto-correlation. 

Once we calculate $G_{k}^{g}(\nu)$ and $G_{k}^{s}(\nu)$, each component (translational, rotational) of the thermodynamic quantities (energy from Eq. \ref{eq-energy} and entropy from Eq. \ref{eq-entropy}) can be determined by integrating the DoS using appropriate weighting functions for the respective thermodynamic quantities:
\begin{equation}
E_{k} = \beta^{-1} \left[\int_{0}^{\infty} d\nu G_{k}^{s}(\nu) W_{E, k}^{s}(\nu) + \int_{0}^{\infty} d\nu G_{k}^{g}(\nu) W_{E, k}^{g}(\nu) \right] ,
\end{equation}
\begin{equation}
S_{k} = k_{B} \left[\int_{0}^{\infty} d\nu G_{k}^{s}(\nu) W_{S, k}^{s}(\nu) + \int_{0}^{\infty} d\nu G_{k}^{g}(\nu) W_{S, k}^{g}(\nu) \right] ,
\end{equation}
\begin{equation}
A_{k} = \beta^{-1} \left[\int_{0}^{\infty} d\nu G_{k}^{s}(\nu) W_{A, k}^{s}(\nu) + \int_{0}^{\infty} d\nu G_{k}^{g}(\nu) W_{A, k}^{g}(\nu) \right] ,
\end{equation}
where, $\beta = (k_{B} T)^{-1}$ and  $W_{l, k}^{g/s}$ is the weighting function for thermodynamic quantity $l$ (E/S/A) for each mode $k$ (translation/rotation) partitioned into gas-like ($g$) or solid-like ($s$) contribution. Here,
\begin{equation}
 W_{E}^{s} = \frac{\beta h \nu}{2} + \frac{\beta h \nu}{exp (\beta h \nu) - 1} ,
\end{equation}
\begin{equation}
 W_{S}^{s} = \frac{\beta h \nu}{exp (\beta h \nu) - 1} - ln[1-exp(-\beta h \nu)] ,
\end{equation}\begin{equation}
 W_{E, trans}^{g}(\nu) = W_{E, rot}^{g} (\nu) = 0.5,
\end{equation}\begin{equation}
 W_{S, trans}^{g}(\nu) = \frac{1}{3}\frac{S^{HS}}{k_{B}},
\end{equation}
\begin{equation}
 W_{S, rot}^{g}(\nu) = \frac{1}{3}\frac{S^{R}}{k_{B}}
\end{equation}where, $S^{HS}$  is the hard-sphere entropy and  $S^{R}$ is the rotational entropy of ideal gas modelled as rigid rotor: 
\begin{equation}
\frac{S^{HS}}{k_{B}} = \frac{5}{2} + ln\left[\left(\frac{2\pi mk_{B}T}{h^{2}}\right)^{3/2} \frac{V}{f_{tr}N}z(y)\right] + \frac{y(3y-4)}{(1-y)^{2}},
\end{equation}
\begin{equation}
\frac{S^{R}}{k_{B}} = 1+ln\left[\frac{T}{\sigma \Theta_{r}} \right],
\end{equation}
here, $y = f_{trans}^{5/2}/ \Delta_{trans}^{3/2}$ and $z(y)$ is the compressibility factor of hard sphere from the Carnahan-Starling equation of state \cite{carnahan1970thermodynamic}. $\Theta_{r} $ is the rotational temperature defined as $\Theta_{r} = \frac{h^{2}}{8\pi^{2}I_{r}k_{B}} $ and $\sigma$ is the rotational symmetry. The reference energy now becomes,
\begin{equation}
E_{0} = E_{MD} - \beta^{-1} 3N(1-0.5f_{trans}-0.5f_{rot}) ,
\end{equation}
where, $E_{MD}$ is the total energy calculated from the MD simulation.

\subsection{Entropy using multiparticle correlation expansion method and integration approach on the SRS equation of state}

The configurational entropy $S_{con}$ is defined as: \cite{costa1998cpl,lazaridis1992entropy,lazaridis1996orientational,cuetos-2002}.
\begin{equation}
    S_{tot}^{con} = S_{id} + \sum_{n=2}^{\infty}S_{n} ,
\end{equation}
where, $S_{id}$ denotes the entropy of an ideal gas and $S_{n}$ denotes the entropy due to n-particle spatial correlation. Therefore, the excess entropy can be calculated from well-known multi-particle correlation expansion of the configurational entropy (ME) $S{ex}$ can be written as:
\begin{equation}
    S_{ex} = \sum_{n = 2}^{\infty}S_{n} = S_{tot}^{con}-S_{id} ,
    \label{eq-s-ex}
\end{equation}
If $S_{2}$ represents the entropy due to pair interaction, then the residual entropy $\Delta s$ that includes the spatial correlation for $n \ge 3$ becomes: 
\begin{equation}
\Delta s = S_{ex}-S_{2} .
\label{eq-residual-entropy}
\end{equation}

Pair entropy $S_{2}$ can be expressed as:
\begin{equation}
    S_{2} = S_{2}^{trans}+S_{2}^{rot} ,
    \label{eq-pair-entropy}
\end{equation}

\begin{equation}
S_{2}^{trans} = -2\pi\rho\int \left[ g(r) lng(r) - g(r) + 1\right] r^{2} dr,  \label{eq-pair-entropy-trans}
\end{equation}

\begin{equation}
S_{2}^{rot} = 4\pi\rho\int g(r)q^{rot}(r)r^{2}dr,
 \label{eq-pair-entropy-rot}
\end{equation}

\begin{equation}
q^{rot} (r) = -\frac{1}{4}\int_{0}^{\pi} g(\theta|r)sin\theta d\theta.
\end{equation}

In a system of linear molecules, the  probability distribution function $g(r, \theta)$ can be factorized as \cite{costa1998cpl},  $g(r, \theta) = g(r) g(\theta|r)$, where, $g(r)$ denotes the radial distribution and $g(\theta|r)$ denotes the conditional probability distribution function  between two rods at a $r$ distance with a relative angle between $\theta$ to $\theta + d\theta$.

The excess entropy can be exactly calculated using the equation of state (EOS) of the SRS defined below\cite{cuetos-2002}:
\begin{equation}
    S^{EOS}_{ex} (\rho) = \frac{U_{ex}}{T} - \int_{0}^{\rho} \left[\frac{P}{k_{B}T\rho'} - 1\right] \frac{ d\rho'}{\rho'},  \label{eq-ex-entropy}
\end{equation}where, $U_{ex}$ represents the excess energy, which is the potential energy per particle in the units of $k_{B}$.


\section{\label{sec:level2}Results and Discussion}
\begin{figure} [!htb]
	\centering
	\includegraphics[scale=0.95]{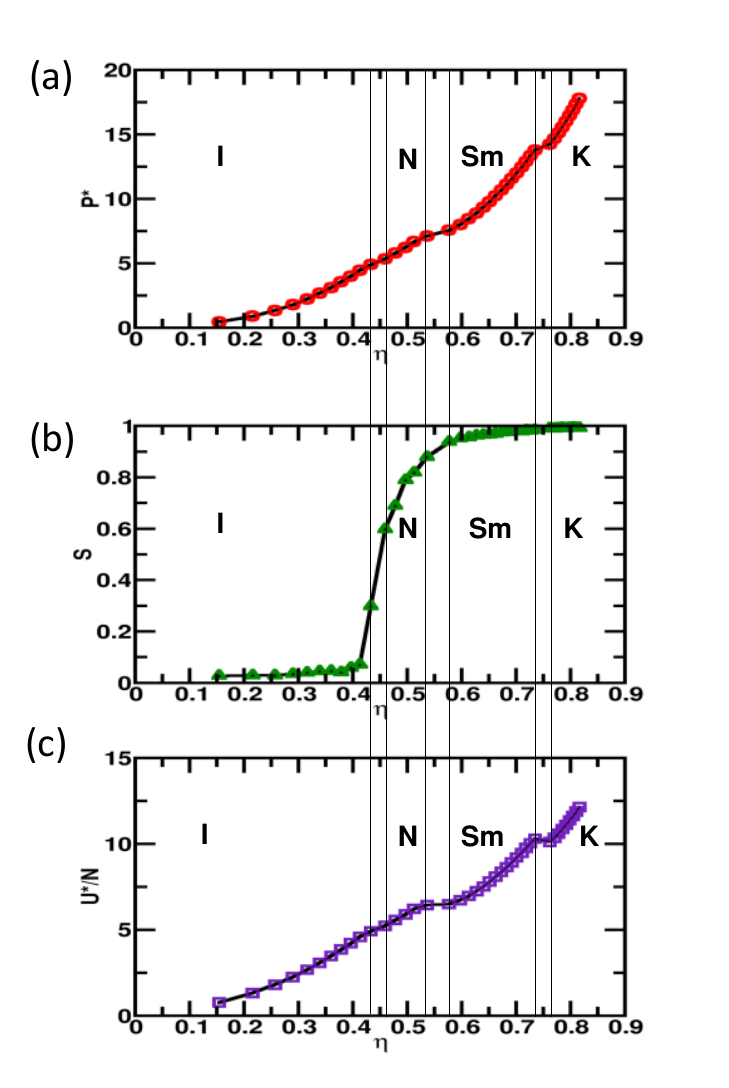}
	\caption{(a) Equation of state (b) nematic order parameter $S$ (c) potential energy per particle $U^{*}/N$ are plotted with packing fraction $\eta$ for the system of soft repulsive spherocylinders of aspect ratio $L/D = 5$. Thermodynamic quantities are defined in the reduced unit: pressure $P^{*} = Pv_{hsc}/k_{B}T$ and packing fraction, $\eta = \rho v_{hsc}$ where $v_{hsc}$ is the volume of the spherocylinder. We observe four stable phases: isotropic (I), nematic (N), smectic (Sm) and crystal (K). The vertical gray lines indicate boundaries between two phases.} \label{eos}
\end{figure}

We present equilibrium phase diagram of SRS of aspect ratios $L/D = 2-5 $ at the temperature $T^{*} = 5 $ (Fig. \ref{eos} and Fig.\ref{eos_s_diffL}(a)). The magnitude of the pressures and densities corresponding to different phases for different aspect ratios are listed in table \ref{table-allL}.
We obtain 4 stable phases for $L/D \ge 3.5: $\cite{cuetos-2002, cuetos-2015,earl2001,Active-SRS,Ons-JC} crystal (K), smectic (Sm), nematic (N) and isotropic (I); 3 stable phases for $L/D = 3 $: crystal, smectic, and isotropic  and two stable phases for $L/D = 2 $: crystal and isotropic. For further details of these phases and their characterization, we refer the reader to our earlier work \cite{Active-SRS,Ons-JC}. Here we are interested in entropy computations of these phases. 

\subsection{Validation of 2PT method}
In the dilute limit, the entropy and Helmholtz free energy of SRS, calculated using 2PT method, can be compared with the values obtained for  an ideal diatomic gas modeled as a rigid rotor. The analytical expressions for the partition function $Z$, entropy $S$, and Helmholtz free energy $A$ of an ideal rigid rotor are as follows: 
\begin{equation}
    Z(V,T) = \left(\dfrac{2\pi mk_{B}T}{h^{2}} \right)^{3/2} V \dfrac{8\pi^{2}Ik_{B}T}{\sigma h^{2}},  \end{equation}
    
\begin{equation}
\dfrac{S}{Nk_{B}} =  ln\left[\frac{2\pi(m_{1}+m_{2})k_{B}T}{h^{2}}  \right]^{3/2} \frac{Ve^{5/2}}{N} + ln \frac{8\pi^{2}Ik_{B}Te}{\sigma h^{2}}.
\label{entropy-idealgas}
\end{equation}  

\begin{equation}
\dfrac{A}{Nk_{B}T} =  -\left(ln\left[\frac{2\pi(m_{1}+m_{2})k_{B}T}{h^{2}}  \right]^{3/2} \frac{V}{N} + ln \frac{8\pi^{2}Ik_{B}T}{\sigma h^{2}} + 1\right).
\label{helmholtz-idealgas}
\end{equation} 

The $1^{st}$ term in Eq. \ref{entropy-idealgas} is due to the translational motion, and the $2^{nd}$ term is due to the rotational motion (for an ideal rigid rotor, there is no vibrational motion). In Table \ref{table-1} and \ref{table-2}, we compare the entropy of the SRS system in a dilute limit calculated from the 2pt method with that of an ideal rigid rotor at the same state point calculated using the above equations for different aspect ratios which are found to be in a very good agreement.

\begin{table}[ht]
	\caption{Comparison of the total $S_{tot}$, translational $S_{trans}$ and rotational $S_{rot}$ entropy 
	of SRS of different aspect ratios from the 2PT method at the temperature $T^{*}=5$ and number density $\rho^{*}=0.01$ with that of a rigid rotor at the same state points calculated using Eq. \ref{entropy-idealgas}. Here, entropy is calculated in $k_{B}/particle$ unit. }
	\vspace{0.4cm}
	\centering
	\begin{tabular}{|c |c |c |c |c |c |c |c|}
		\hline
		
		$L/D$ & $\rho^{*}$ & $S_{trans}^{id}$ & $S_{rot}^{id}$ & $S_{tot}^{id}$ & $S_{trans}^{2PT}$ & $S_{rot}^{2PT}$ & $S_{tot}^{2PT}$ \\[2ex]
		\hline
		5 & 0.01 & 18.36 & 12.18 & \textbf{30.54} & 18.26 & 12.30 & \textbf{30.56} \\[2ex]
		\hline
		3 & 0.01 & 18.36 & 11.15 & \textbf{29.51} & 18.36 & 11.25 & \textbf{29.61} \\[2ex]
		\hline
		2 & 0.01 & 18.36 & 10.34 & \textbf{28.70} & 18.36 & 10.44 & \textbf{28.80} \\[2ex]
		\hline
		
	\end{tabular}
	\label{table-1}
\end{table}
\begin{table}[ht]
	\caption{Comparison of the Helmholtz free energy of SRS of different aspect ratios from the 2PT method with that of the ideal rigid rotor using Eq.\ref{helmholtz-idealgas} at the dilute limit, temperature $T^{*}=5$ and number density $\rho^{*}=0.01$. $A^{*}_{tot}$ designates the total Helmholtz free energy and $A^{*}_{trans}$, $A^{*}_{rot}$ designate the translational and rotational components respectively.}
	\vspace{0.4cm}
	\centering
	\begin{tabular}{|c |c |c |c |c |c |c |c|}
		\hline
		
		$L/D$ & $\rho^{*}$ & $A_{trans}^{id}$ & $A_{rot}^{id}$ & $A_{tot}^{id}$ & $A_{trans}^{2PT}$ & $A_{rot}^{2PT}$ & $A_{tot}^{2PT}$ \\[2ex]
		\hline
		5 & 0.01 & -84.24 & -55.81 & \textbf{-139.95} & -84.91 &-55.95 & \textbf{-140.86} \\[2ex]
		\hline
		3 & 0.01 & -84.24 & -50.71 & \textbf{-134.98} & -85.63 & -50.65 & \textbf{-136.28} \\[2ex]
		\hline
		2 & 0.01 & -84.24 & -46.66 & \textbf{-130.88} & -83.74 & -48.33 & \textbf{-132.07} \\[2ex]
		\hline
		
	\end{tabular}
	\label{table-2}
\end{table}

\subsection{Density of states of liquid crystal phases}
\begin{figure*} [!htb]
	\centering
	\includegraphics[scale=0.80]{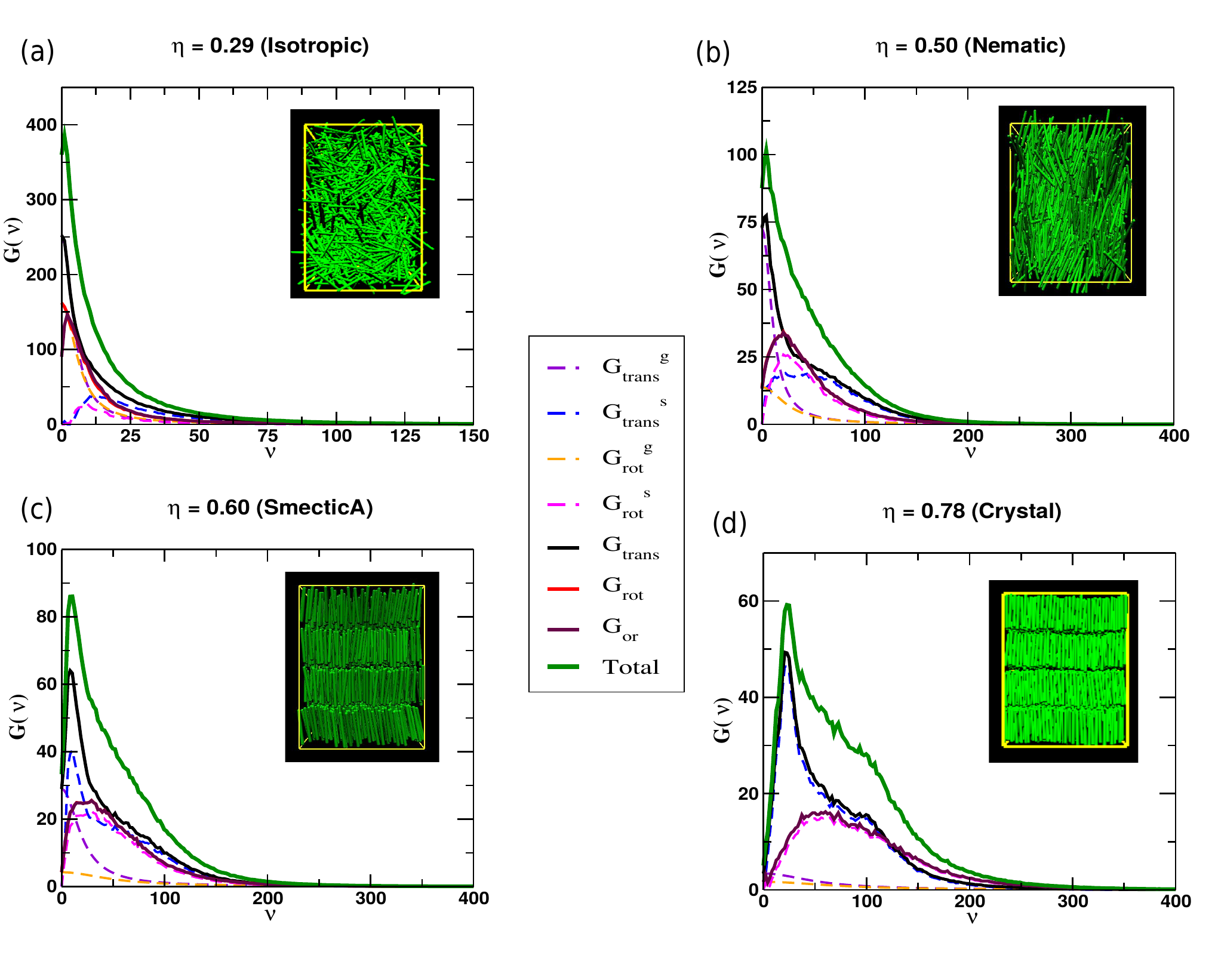}
	\caption{Density of state (DoS) $G(\nu)$ for (a) isotropic (b) nematic (c) smectic and (d) crystal phase. The components of the entropy are mentioned in the legend. The snapshots of the configurations are shown for the respective phases. Here, we see that DoS of nematic phase [Fig. (b)] comprises both solid and gas like components, whereas for smectic phase [Fig.(c)], it is dominated by solid like components only.} \label{DoS}
\end{figure*}

We calculate the density of state $G(\nu)$ of different liquid crystal phases using 2PT method as shown in Fig. \ref{DoS}. For each phase, we show the total DoS and its decomposition into translational, rotational modes. The translational and rotational modes are further decomposed into gas-like and solid-like components, as mentioned in the 2PT method section. \\

In Fig.\ref{DoS}(a), we plot DoS for the state point $P^{*} = 1.78, \eta = 0.29 $ which corresponds to the isotropic phase as shown in the equilibrium phase diagram [Fig.\ref{eos}]. We find that both the translational $G_{trans}$ and rotational $G_{rot}$ DoS are dominated by the gas like contribution and decay exponentially. At the zero frequency $\nu = 0 $, both of $G_{trans}$ and $G_{rot}$ have large finite values, indicating that the system possesses high translational and rotational diffusivity. 
Similarly, in Fig.\ref{DoS}(b), we plot DoS of nematic phase for the state point $P^{*} = 6.23, \eta = 0.50 $. We see that $G_{trans}$ decays exponentially and have a fine value at $\nu = 0$ indicating gas-like behaviour. However, $G_{rot}$ is dominated by solid-like behaviour with a low rotational diffusivity. 
In the case of the smectic phase ( $P^{*} = 8, \eta = 0.6 $), both of the $G_{trans}$ and $G_{rot}$ are dominated by solid-like contribution. However, $G_{trans}$ has a very low value at zero-frequency indicating a low-diffusivity which is due to the in-layer fluid-like motion. 
In crystal phase (state point $P^{*} = 15.13, \eta = 0.78 $), $G(\nu)$ is roughly zero at $\nu = 0$ indicating absence of diffusive mode in the system. Both translational and rotational DoS exhibit solid-like behaviour.

\subsection{Fluidicity factor of liquid crystal phases}
\begin{figure} [!htb]
	\centering
	\includegraphics[scale=0.48]{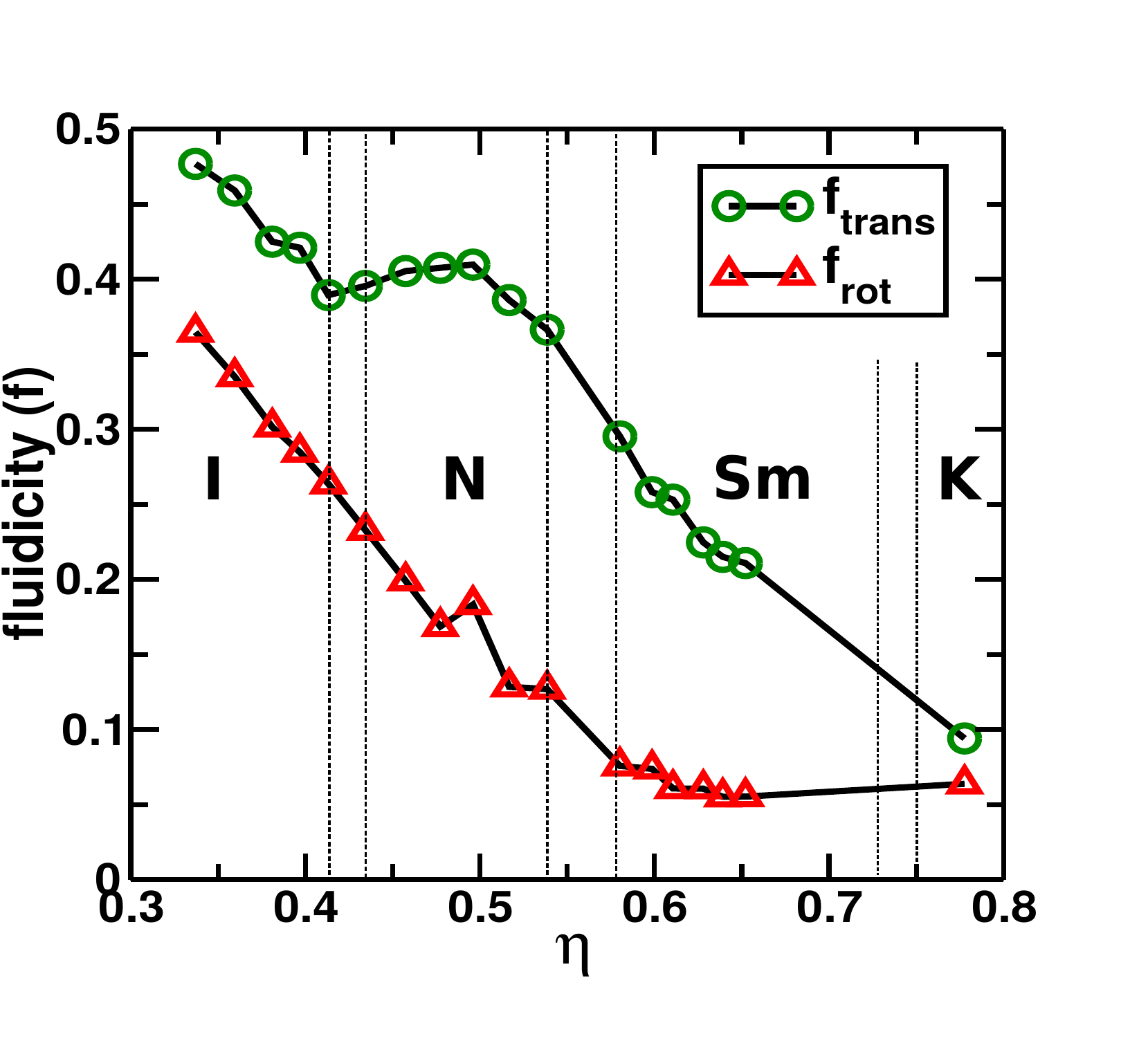}
	\caption{Phase diagram of the SRS with aspect ratio $L/D = 5$ at $T^{*} = 5$ in fluidicity, packing fraction ($f-\eta$) space. $f_{trans}$ and $f_{rot}$ represent the translational and rotational components respectively.
	The black dotted lines denote phase boundaries of different phases. 
	} \label{fluidicity}
\end{figure}

The decomposition of the translational and rotational DoS into gas-like and solid-like components is carried out by calculating the fluidicity factor $f$ as discussed in Section III-A. We find that both of translational and rotational fluidicity factors are very high in the isotropic phase, very low in the crystal phase and intermediate in the LC phases as mentioned in the Table \ref{table-f} and in Fig. \ref{fluidicity}. 
We also calculate the phase boundaries of different LC phases from the change of $f_{trans}$ and $f_{rot}$.  In Fig.\ref{fluidicity}, we find that, both of the $f_{trans}$ and $f_{rot}$ decrease with packing fraction $\eta$ in the isotropic phase. In the nematic phase, $f_{trans}$ remains almost constant at its value in the isotropic phase, while $f_{rot}$ keeps decreasing. This is also consistent with the DoS calculation showing that rotational diffusivity is much lower in the nematic phase than translational diffusivity. The I-N phase boundary is therefore defined as the packing fraction where $f_{rot}$ keeps decreasing but $f_{trans}$ becomes constant ($\eta^{*}_{I-N} \approx 0.41-0.44 $ for $L/D = 5$). 
Similarly, in the Smectic phase, $f_{rot}$ remains nearly constant at its value in the nematic phase while $f_{trans}$ drops sharply. Hence, the N-Sm phase boundary can be located at the packing fraction where $f_{trans}$ continues to decrease but $f_{rot}$ remains almost constant ($\eta^{*}_{N-Sm} \approx 0.54-0.57 $ for $L/D = 5$). Both of $f_{trans}$ and $f_{rot}$ acquire a very low value in the crystal phase. These analyses suggest another method of quantifying the phase boundaries using the fluidicity factor.


\begin{table}[ht]
	\caption{Translational and rotational fluidicity factors for different liquid crystal phases for the aspect ratio $L/D = 5$.}
	\vspace{0.4cm}
	\centering
	\begin{tabular}{|c |c |c |c |c |}
 
		\hline
            $P^{*}$ & $\eta$ & $f_{trans}$ & $f_{rot}$ & Phase \\[2ex]
            \hline
            1.78 & 0.29 & 0.62 & 0.53 & I \\[2ex]
		\hline
		  6.23 & 0.50 & 0.41 & 0.18 & N  \\[2ex]
		\hline
		  8.01 & 0.60 & 0.26 & 0.07 & Sm  \\[2ex]
		\hline
		15.13 & 0.78 & 0.09 & 0.06 & K \\[2ex]
		\hline
		
	\end{tabular}
	\label{table-f}
\end{table}

\subsection{Entropy calculation from 2PT method}
\begin{figure} [!htb]
	\centering
	\includegraphics[scale=0.43]{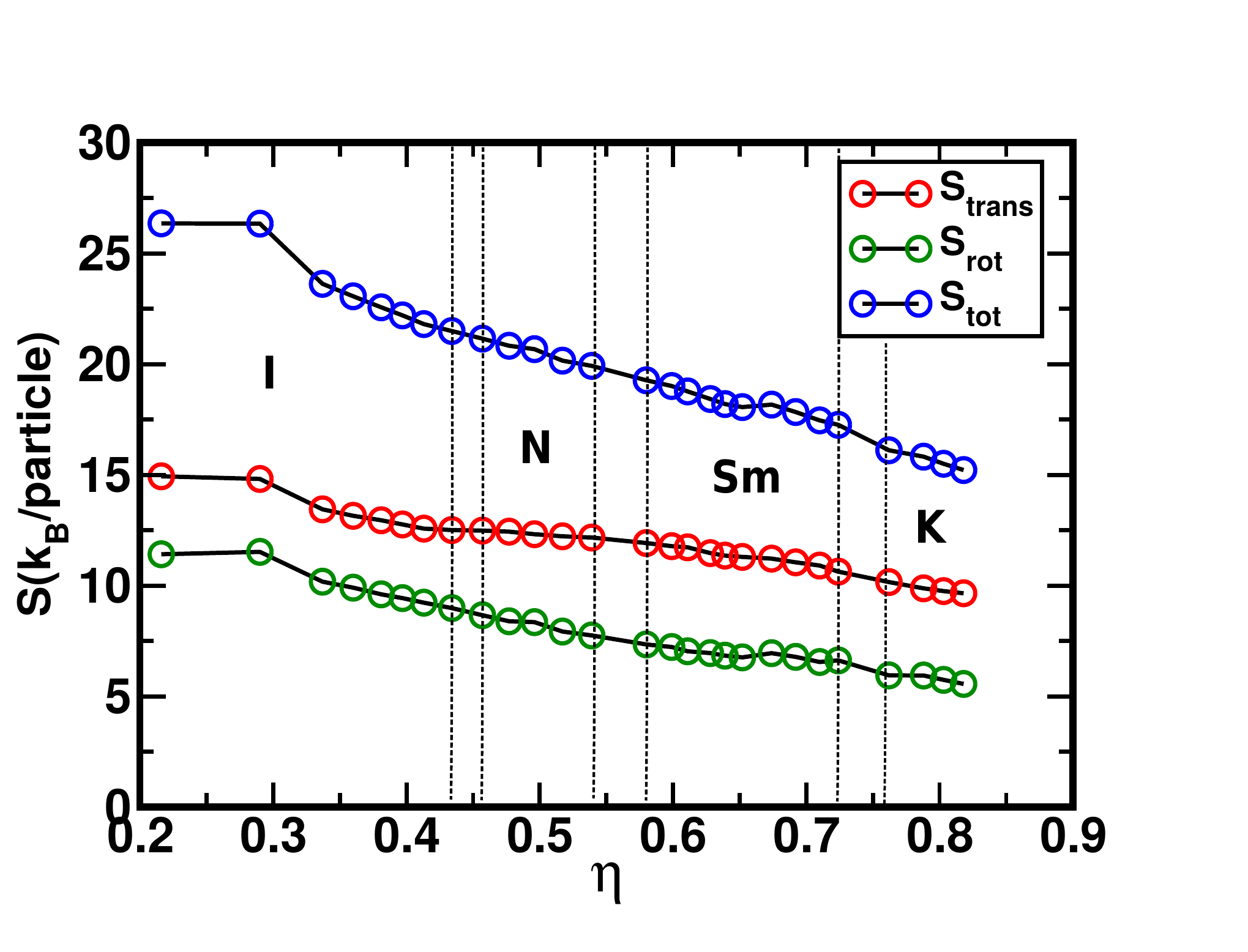}
	\caption{Total entropy and its translational and rotational components of different liquid crystal phases for the aspect ratio $L/D = 5$ at $T^{*}=5$. The black dotted lines denote the phase boundaries.} \label{entropy}
\end{figure}

\begin{figure*} [!htb]
	\centering
	\includegraphics[scale=1]{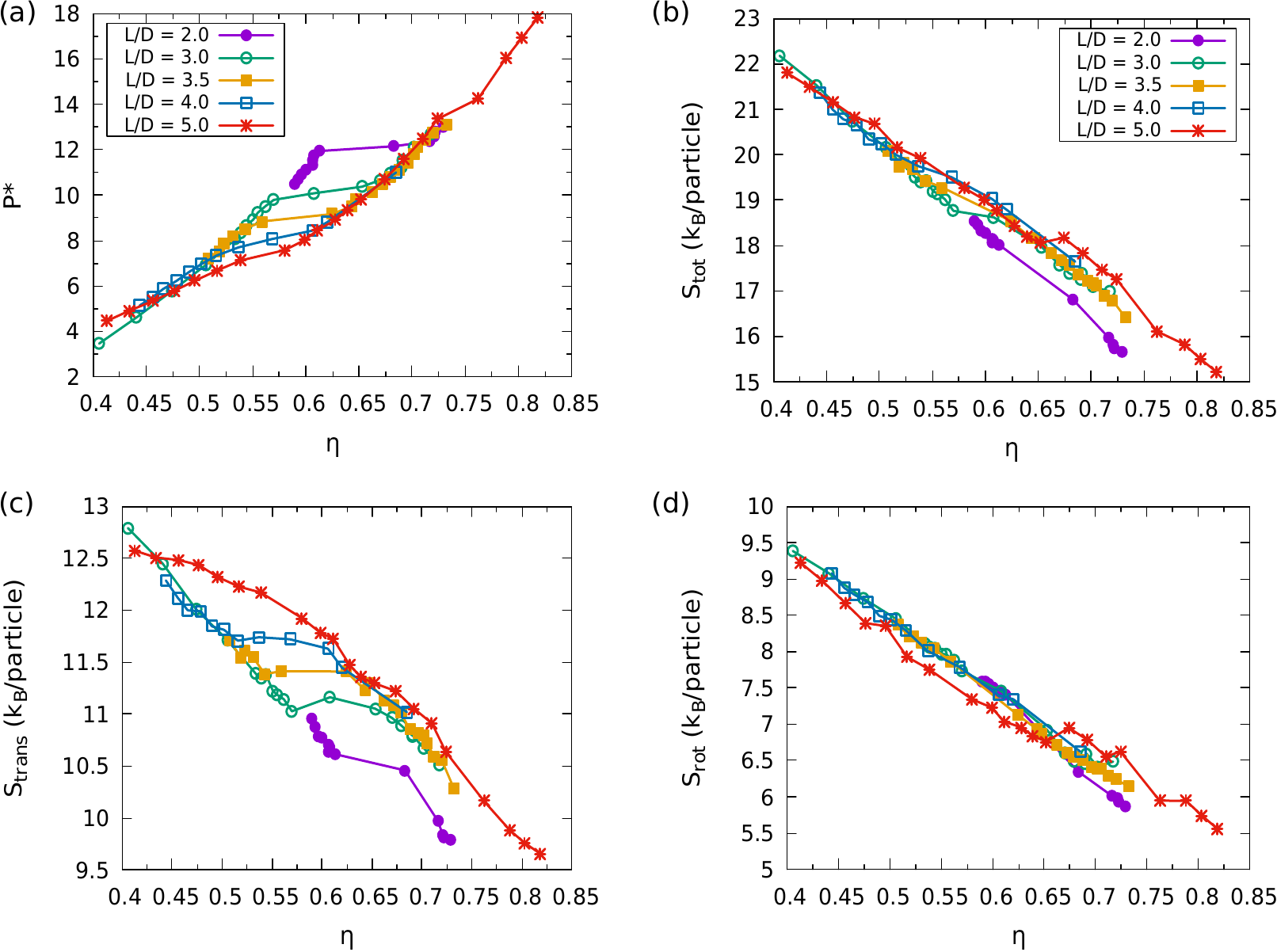}
	\caption{(a) Equation of state, (b) total entropy $S_{tot}$ and its decomposition into (c) translational $S_{trans}$ and (d) rotational $S_{rot}$ motion as a function of packing fraction $\eta$ for different $L/D $s at the temperature $T^{*} = 5$. Here we see that, at a certain packing fraction, total entropy is roughly same irrespective of the different LC phases corresponding to different L/Ds.} \label{eos_s_diffL}
\end{figure*}

In Table \ref{table-allL} and in Fig.\ref{entropy}, Fig.\ref{eos_s_diffL}, we mention the total entropy $S_{tot}$ and its decomposition into the translational $S_{trans}$ and rotational $S_{rot}$ modes for different liquid crystal phases associated to different aspect ratios. 
We find that entropy decreases as a function of packing fraction for the given aspect ratios. 
We also find that, at a certain packing fraction, total entropy is close by for the given aspect ratios, irrespective of the different liquid crystal phases they exhibit.
As for example, at $\eta = 0.60$, the magnitude of the total entropy is $S_{tot} = 18.54-19.03$ $k_{B}$/particle; however, it shows smectic structure for $L/D \geq 3 $ and isotropic structure for $L/D = 2 $. Similarly, at $\eta = 0.54$, $S_{tot} = 19.40-19.92$ $k_{B}$/particle while it shows nematic structure for $L/D \geq 3.5 $ and isotropic structure for $L/D = 3, 2 $. These results indicate that, total entropy depends on the thermodynamic state points only, not on the different liquid crystal phases corresponding to different $L/D$ s. However, the entropy of different $L/D$ s differs at the higher packing fractions, as mentioned in Fig. \ref{eos_s_diffL}(b). 

\begin{figure} [!htb]
	\centering
	\includegraphics[scale=1.1]{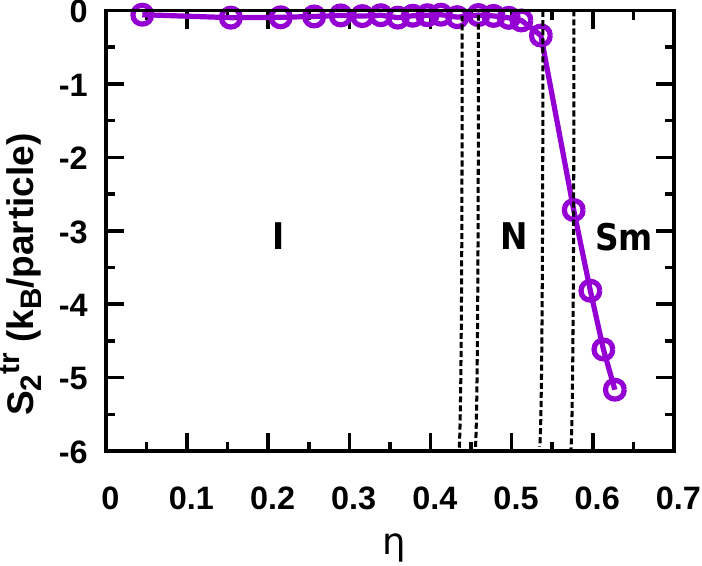}
	\caption{Translational pair entropy per particle of different liquid crystal phases for the aspect ratio $L/D = 5$ at $T^{*}=5$. The black dotted lines denote the phase boundaries.} \label{tr-pair-entropy}
\end{figure}

In Fig.\ref{tr-pair-entropy}, we calculate the pair entropy $S_{2}$ of different LC phases using Eq. \ref{eq-pair-entropy} and its decomposition into translational $S_{2}^{tr}$ and rotational $S_{2}^{rot}$ parts. We observe that, $S_{2}^{rot}$ decreases sharply at the I-N phase boundary, while $S_{2}^{tr}$ decreases slowly. For N-Sm transition, $S_{2}^{tr}$ decreases more rapidly than that of I-N phase boundary. Our results are consistent with those of Cuetos \textit{et al.} \cite{cuetos-2002}. 
These analyses indicate that the change of entropy at the LC phase transition points are mainly driven by the translational or rotational pair entropy. The sharp decrease of rotational pair entropy at the I-N phase boundary is compensated by residual entropy $\Delta s$ arising from the multi particle correlation (Eq. \ref{eq-residual-entropy}). Similarly, the N-Sm phase transition is driven by the sharp decrease of translational pair entropy that is also compensated by residual entropy.

\begin{figure} [!htb]
	\centering
	\includegraphics[scale=0.5]{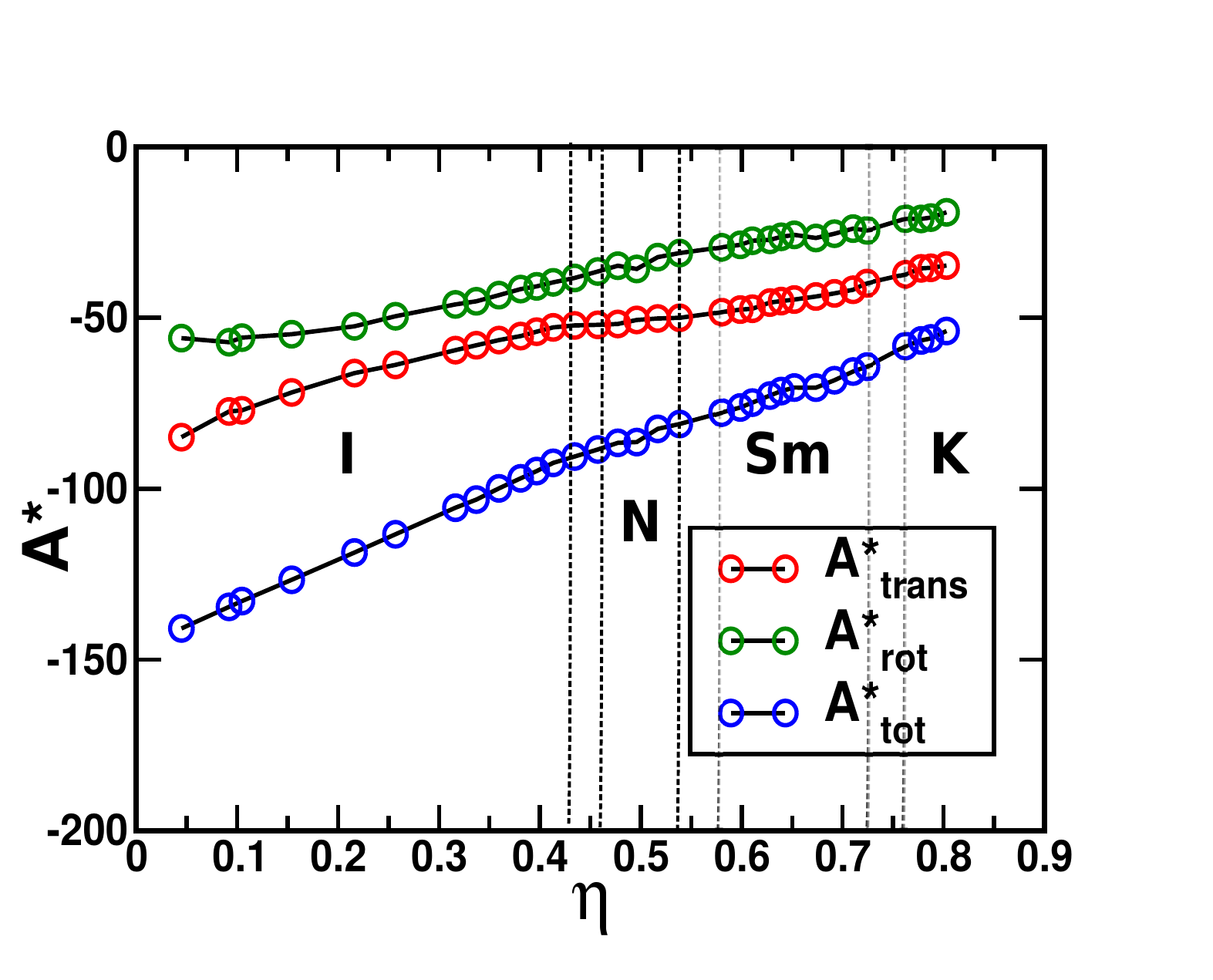}
	\caption{Helmholtz free energy $A^*_{tot}$ and its translational $A^*_{trans}$ and rotational $A^*_{rot}$ components of different liquid crystal phases for the aspect ratio $L/D = 5$ at $T^{*}=5$. The black dotted lines denote the phase boundaries.} \label{free-energy}
\end{figure}

\subsection{Comparison of excess entropy from 2PT method and integrating on the SRS equation of state}

\begin{figure*} [!htb]
	\centering
	\includegraphics[scale=0.6]{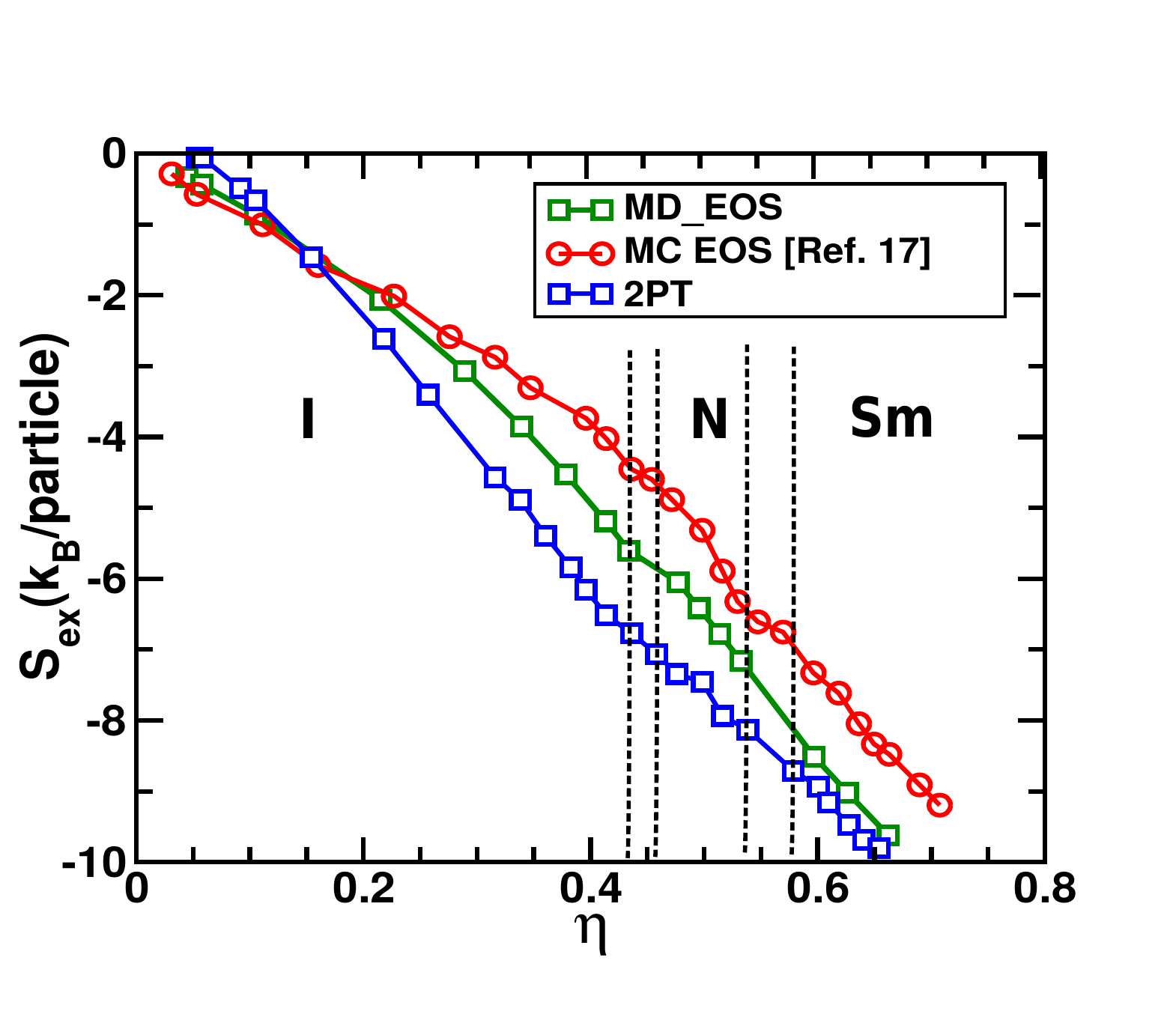}
	\caption{Excess entropy $S_{ex} = S^{2PT/EOS}_{tot}-S^{id}_{tot}$ vs packing fraction $\eta$ for $L/D = 5$ calculated using 2PT method and equation of state (EOS) of SRS. Here, we compare the excess entropy calculated using MD equation of state and 2PT method from our simulation with those of the Monte Carlo (MC) EOS from Cuetos \textit{et al.}\cite{cuetos-2002}. The black dotted lines denote the phase boundaries.}  \label{s_ex}
\end{figure*}

We calculate the excess entropy, $S_{ex}$ which is defined as the amount of entropy arises due to the particles' interaction using Eq.\ref{eq-s-ex}. It is calculated from the difference between the absolute entropy calculated from the 2PT method or integrating over MD/MC equation of state and the entropy of an ideal rigid rotor at the same state point. We mention the magnitude of $S_{ex}$ for different liquid crystal phases in Table \ref{table-sex} for $L/D = 5 $ at $T^{*} = 5$. 
In Fig. \ref{s_ex}, we compare the excess entropy of SRS at different packing fractions from the 2PT method with those of the standard integration approach on the (a) MD equation of state of SRS from our simulation and (b) MC equation of state of SRS employed by Cuetos \textit{et al.} \cite{cuetos-2002} 
We observe that the magnitude of $S_{ex}$ are in good agreement at the lower densities for the given methods. At the higher densities, $S_{ex}$ calculated from 2PT method matches well with the MD equation of state, but it differs from the MC equation of state \cite{cuetos-2002}. 

\begin{table*}[ht]
	\caption{ Total entropy $S_{tot}^{2PT}$ and its decomposition into translational $S_{trans}^{2PT}$ and rotational $S_{rot}^{2PT}$ degrees of freedom for different liquid crystal phases associated to different aspect ratios $L/D $ s at $T^{*} = 5$. Here, $P^{*}, \rho^{*}$ and $\eta$ indicate pressure, number density and packing fraction respectively.}
	\vspace{0.4cm}
	\centering
	\begin{ruledtabular}
	\begin{tabular}{c c c c c c c}
	$P^{*}$ & $\rho^{*}$ & $\eta$ & $S_{trans}^{2PT}$ & $S_{rot}^{2PT}$ & $S_{tot}^{2PT}$ & Phase \\[2ex]
	\hline
	& & &  & & & \\[1ex]
	& & &  L/D = 5 & & & \\[2ex]
	\hline
    4.45 & 0.093 & 0.413 & 12.574 & 9.230 & 21.804  & I \\[2ex]
    4.90 & 0.098 & 0.434 & 12.509 & 8.985 & 21.494 & N \\[2ex]
    5.34 & 0.103 & 0.457 & 12.483 & 8.667 & 21.150 & N \\[2ex]
    5.79 & 0.107 & 0.477 & 12.435 & 8.391 & 20.827 & N \\[2ex]
    6.23 & 0.111 & 0.496 & 12.320 & 8.354 & 20.674 & N \\[2ex]
    6.68 & 0.116 & 0.517 & 12.228 & 7.928 & 20.157 & N \\[2ex]
    7.12 & 0.121 & 0.539 & 12.170 & 7.751 & 19.921 & N \\[2ex]
    7.57 & 0.130 & 0.580 & 11.924 & 7.344 & 19.268 & SmA \\[2ex]
    8.01 & 0.135 & 0.599 & 11.780 & 7.231 & 19.011 & SmA \\[2ex]
    8.46 & 0.137 & 0.611 & 11.730 & 7.039 & 18.768 & SmA \\[2ex]
    9.35 & 0.144 & 0.639 & 11.355 & 6.841 & 18.197 & SmA \\[2ex]
    9.79 & 0.147 & 0.652 & 11.298 & 6.761 & 18.059 & SmA \\[2ex]
    10.68 & 0.151 & 0.674 & 11.221 & 6.951 & 18.172 & SmA \\[2ex]
    12.46 & 0.160 & 0.710 & 10.909 & 6.553 & 17.462 & SmA \\[2ex]
    14.24 & 0.171 & 0.762 & 10.161 & 5.950 & 16.111 & SmA \\[2ex]
    16.02 & 0.177 & 0.788 & 9.879  & 5.944 & 15.823 & K \\[2ex]
    17.80 & 0.184 & 0.818 & 9.656  & 5.563 & 15.219 & K \\[2ex]

	\hline
	& & &  & & & \\[1ex]
	& & &  L/D = 4 & & & \\[2ex]
	\hline

    5.13 & 0.121 & 0.444 & 12.287 & 9.081 & 21.368 & I \\[2ex]
    5.86 & 0.127 & 0.466 & 11.999 & 8.783 & 20.782 & I \\[2ex]
    6.60 & 0.134 & 0.490 & 11.845 & 8.493 & 20.338 & I \\[2ex]
    6.96 & 0.137 & 0.502 & 11.812 & 8.438 & 20.251 & N \\[2ex]
    7.33 & 0.141 & 0.516 & 11.706 & 8.290 & 19.996 & N \\[2ex]
    7.70 & 0.146 & 0.537 & 11.740 & 8.008 & 19.748 & N \\[2ex]
    8.06 & 0.155 & 0.568 & 11.720 & 7.787 & 19.507 & SmA \\[2ex]
    8.43 & 0.165 & 0.606 & 11.630 & 7.401 & 19.031 & SmA \\[2ex]
    8.80 & 0.169 & 0.620 & 11.446 & 7.346 & 18.792 & SmA \\[2ex]
    11.00 & 0.187 & 0.685 & 11.019 & 6.625 & 17.644 & K \\[2ex]
	\end{tabular}
	\label{table-allL}
	\end{ruledtabular}
\end{table*}
\begin{table*}[ht]
	\vspace{0.4cm}
	\centering
	\begin{ruledtabular}
	\begin{tabular}{c c c c c c c}
	$P^{*}$ & $\rho^{*}$ & $\eta$ & $S_{trans}^{2PT}$ & $S_{rot}^{2PT}$ & $S_{tot}^{2PT}$ & Phase \\[2ex]
    \hline
	& & &  & & & \\[1ex]
	& & &  L/D = 3.5 & & & \\[2ex]
	\hline

   7.20  & 0.155  & 0.508  & 11.708  & 8.369  & 20.077 & I \\[2ex]
   7.85  & 0.160  & 0.523  & 11.608  & 8.203  & 19.810 & I \\[2ex]
   8.18  & 0.162  & 0.531  & 11.549  & 8.135  & 19.684 & N \\[2ex]
  8.51  & 0.166  & 0.543  & 11.385  & 8.041  & 19.427 & N \\[2ex]
   8.84  & 0.171  & 0.559  & 11.412  & 7.873  & 19.285 & N \\[2ex]
   9.16  & 0.191  & 0.624  & 11.410  & 7.134  & 18.544 & SmA \\[2ex]
  9.49  & 0.196  & 0.643  & 11.233  & 6.931  & 18.164 & SmA \\[2ex]
   9.82  & 0.198  & 0.647  & 11.306  & 6.870  & 18.176 & SmA \\[2ex]
   10.14 &  0.203 &  0.663 &  11.130 &  6.714 &  17.844 & K \\[2ex]
   10.47 &  0.205 &  0.672 &  11.082 &  6.607 &  17.688 & K \\[2ex]
   13.09 &  0.224 &  0.732 &  10.282 &  6.142 &  16.423 & K \\[2ex]

    \hline
	& & &  & & & \\[1ex]
	& & &  L/D = 3 & & & \\[2ex]
	\hline
2.30 & 0.122 & 0.352 & 13.438 & 9.802 & 23.239 & I                 \\[2ex]
6.91 & 0.176 & 0.506 & 11.708 & 8.453 & 20.161 & I                \\[2ex]
8.06 & 0.185 & 0.534 & 11.387 & 8.116 & 19.504 & I                 \\[2ex]
8.35 & 0.187 & 0.539 & 11.345 & 8.058 & 19.403 & I                \\[2ex]
9.50 & 0.195 & 0.562 & 11.142 & 7.878 & 19.020 & I                 \\[2ex]
9.79 & 0.198 & 0.570 & 11.026 & 7.736 & 18.762 & SmA               \\[2ex]
10.08 & 0.211 & 0.608 & 11.161 & 7.460 & 18.621 & SmA             \\[2ex]
10.37 & 0.227 & 0.653 & 11.052 & 6.920 & 17.972 & SmA              \\[2ex]
10.66 & 0.233 & 0.670 & 10.963 & 6.614 & 17.577 & K                \\[2ex]
12.67 & 0.249 & 0.717 & 10.507 & 6.486 & 16.992 & K                \\[2ex]

   \hline
	& & &  & & & \\[1ex]
	& & &  L/D = 2 & & & \\[2ex]
	\hline
10.47 & 0.282 & 0.590 & 10.955 &  7.584 &  18.539 & I \\[2ex]
11.10 & 0.287 & 0.600 & 10.776 &  7.510 &  18.286 & I \\[2ex]
11.94 & 0.293 & 0.613 & 10.614 &  7.404 &  18.017 & I \\[2ex]
12.15 & 0.326 & 0.683 & 10.457 &  6.351 &  16.808 & K \\[2ex]
12.57 & 0.344 & 0.721 & 9.830  & 5.992  & 15.822 & K \\ [2ex]
12.99 & 0.348 & 0.729 & 9.786  & 5.862  & 15.647 & K \\ [2ex]
    \end{tabular}
	\end{ruledtabular}
\end{table*}

\begin{table*}[ht]
	\caption{Total entropy $S_{tot}^{2PT}$ and the excess entropy $S^{2PT}_{ex}$ calculated from the 2PT method, entropy of ideal rigid rotor $S_{tot}^{id}$ calculated using Eq.  \ref{entropy-idealgas} and excess entropy using Monte Carlo equation of state from Cuetos \textit{et al.}\cite{cuetos-2002}  $S^{EOS}_{ex}$ for different liquid crystal phases of $L/D = 5$ at $T^{*} = 5$:}
	\vspace{0.4cm}
	\centering
	\begin{ruledtabular}
	\begin{tabular}{c c c c c c c}
		$P^{*}$ & $\rho^{*}$ & $S_{tot}^{2PT}$ & $S_{tot}^{id}$ &$S_{ex}^{2PT}$ &$S_{ex}^{EOS}$ Ref\cite{cuetos-2002} & Phase \\[2ex]
		\hline
		4.45 & 0.093 & 21.803 & 28.316 & -6.512 & -4.023 & I \\[2ex]
		4.90 & 0.098 & 21.494 & 28.267 & -6.772 & -4.454 & N \\[2ex]
		5.34 & 0.103 & 21.150 & 28.215 & -7.064 & -4.598 & N \\[2ex]
		5.79 & 0.107 & 20.827 & 28.172 & -7.345 & -4.885 & N \\[2ex]
        6.23 & 0.112 & 20.674 & 28.134 & -7.459 & -5.316 & N \\[2ex]
        6.68 & 0.116 & 20.156 & 28.093 & -7.936 & -5.891 & N  \\[2ex]
        7.12 & 0.121 & 19.921 & 28.051 & -8.131 & -6.322 & N \\[2ex]
        7.57 & 0.130 & 19.268 & 27.977 & -8.709 & -6.753 & SmA \\[2ex]
        8.01 & 0.135 & 19.011 & 27.946 & -8.935 & -7.328 & SmA \\[2ex]
        8.46 & 0.137 & 18.768 & 27.926 & -9.158 & -7.615 & SmA \\[2ex]
        8.90 & 0.141 & 18.425 & 27.898 & -9.473 & - & SmA \\[2ex]
        9.35 & 0.144 & 18.197 & 27.880 & -9.683 & -8.046 & SmA \\[2ex]
        9.79 & 0.147 & 18.059 & 27.860 & -9.801 & -8.333 & SmA \\[2ex]
	\end{tabular}
	\label{table-sex}
	\end{ruledtabular}
\end{table*}

\section{Conclusion and outlook}
We describe a technique based on the two-phase thermodynamic model (2PT) for computing the entropy of liquid crystal phases of SRS with a range of aspect ratios $L/D = 2-5$. For various liquid crystal phases, we compute the density of state (DoS) functions and 
its decomposition into translational and rotational motions. In the dilute limit, the entropy calculated using the 2PT method matches exactly with that of an ideal rigid rotor. We find that, at a definite packing fraction, the magnitude of the total entropy is roughly equal regardless of the different LC phases associated to different aspect ratios. We compare the excess entropy with that of the conventional integration approach on equation of state of SRS, that matches well. The phase boundaries of different liquid crystal phases are also calculated using the rotational and translational fluidicity factors. Our future study will involve to utilise this method in calculating absolute value of entropy and other thermodynamic quantities of various liquid crystal molecules and compare it with experiments.

\begin{acknowledgments}
	We thank SERB, India for financial support through providing computational facility. JC acknowledges support through an INSPIRE fellowship. JC thanks S. Siva Nasarayya Chari for insightful discussions.
\end{acknowledgments}

\bibliography{aipsamp}



\end{document}